\documentstyle[twocolumn,prl,aps]{revtex}  

\newcommand{\erf}{\mbox{erf}}
\newcommand{\erfc}{\mbox{erfc}}
\newcommand{\Ber}{\mbox{Ber}}
\newcommand{\Bei}{\mbox{Bei}}
\newcommand{\Ker}{\mbox{Ker}}
\newcommand{\Kei}{\mbox{Kei}}

\begin{document}
\draft
\title{Local entropic effects of polymers grafted to soft interfaces.}
\author{T. Bickel\thanks{Author to whom correspondence should be addressed. 
{\em $\mbox{E-mail address}$}: bickel@ldfc.u-strasbg.fr} and C.M. Marques\\}
\address{L.D.F.C.- UMR 7506, 3 rue de l'Universit\'e, 67084 Strasbourg Cedex, France}
\author{C. Jeppesen\\}
\address{ Materials Research Laboratory, 
University of California, Santa Barbara, CA 93106, USA}
\date{April 21, 2000}
\maketitle
\vskip -1truecm
\begin{abstract}
{\bf In this paper, we study the equilibrium properties of polymer chains end-tethered to a fluid
membrane. The loss of conformational entropy of the polymer results in an inhomogeneous pressure
field that we calculate for gaussian chains. We estimate the effects of excluded volume through
a relation between pressure and concentration. Under the polymer pressure, a soft surface will
deform. We calculate the deformation profile for a fluid membrane and show that close to the grafting
point, this profile assumes a cone-like shape, independently of the boundary conditions.
Interactions between different polymers are also mediated by the membrane deformation. This
pair-additive potential is attractive for chains grafted on the same side of the membrane and
repulsive otherwise.} 

\end{abstract}

\narrowtext

\section{Introduction.}
\label{intro}
Fluid membranes are surfactant bilayers self-assembled from solution~\cite{israelbook}. They
 are the prevalent constituents of many natural and
industrial colloidal suspensions, that often contain also other macromolecular
species. In the biological realm, phospholipid bilayers build the walls of liposomes
and cells,  hosting  proteins responsible for functions as diverse as  anchoring  the
cytoskeleton, providing coating protection against the body immune response or 
opening ionic channels for osmotic compensation~\cite{albertsbook}. 
In cosmetics, pharmaceutics or detergency, many formulations are    membrane 
solutions with polymers added  for  performance, processing, conditioning or
delivery~\cite{vandepas}. 

The interactions between polymers and fluid bilayers have
been well scrutinized in many systems. Polymers grafted to the bilayers can induce
gelation~\cite{warriner} or other phase changes~\cite{yiyang,bouglet} in liquid
lamellar phases. They stabilize monodisperse vesicles~\cite{joannic} and modify
the geometry of monolamellar~\cite{ringsdorf,cates} and multilamellar
cylindrical vesicles~\cite{frette}. Theoretically, the behaviour of fluid
membranes is well understood in terms of  bending elasticity~\cite{helfrich1,safranbook}, a description
that requires as an input the value of three material constants: the bending rigidity $\kappa$,
the gaussian rigidity $\bar \kappa$ and the spontaneous curvature radius $R_0$. One might hope
that the behaviour of mixed systems can still be described by an effective elastic energy, with
modified material constants. The task that theoretical studies have undertaken is to calculate the
modifications induced on 
 $\kappa$, $\bar \kappa$ and $R_0$ by the addition of the
macromolecules~\cite{cantor,milner,podgornik,lipowsky1,hanke}.

However, polymer-membrane interactions must have a local quality.
For instance, if a polymer is end-tethered to a membrane, it is clear that the
interactions are strong at the anchoring point and vanish far enough
from it.  We show in this paper that, for grafted
polymers, it is possible to construct a local description of polymer membrane interactions. Our
description stands on the recognition that an end-grafted polymer applies a pressure field to 
the grafting wall~\cite{bickel2,breidenich}. The pressure field results into a local deformation
of the membrane: a  membrane with grafted polymers can therefore be seen as a surface
with bending elasticity carrying a number of pressure patches, each of them
creating its own deformation.

The paper is organized as follows. In the next section we compute the pressure
field applied by a grafted polymer in theta and good solvent conditions. In
section~\ref{sec:5}, we consider the case of a freely-standing membrane, for which we
compute the deformation induced by the polymer pressure patch. We also show in this section
that the interactions between the different deformation fields give rise to a membrane
mediated potential between different grafted polymers. The consequences of such potential
are briefly discussed. Section~\ref{sec:9} is dedicated to two membrane geometries relevant
for experiments. We first discuss the case of supported membranes,  and then the case of
lamellar phases.
In the conclusions, we will briefly speculate on the implications of our
results for hairy vesicles.

\section{The pressure applied by a grafted polymer.}
\label{sec:1}
The number of available conformations for a long, flexible
polymer is strongly reduced by the process of end-grafting the
chain to a hard wall~\cite{degennesbook,eisenbook}. The average configuration of the
macromolecule is a compromise between the need to avoid the
surface and the constraint imposed by the tethered end. It is
clear that if the surface can be deformed, there will be an
entropic reason to push it away from the monomer cloud. This can
be described as a pressure that the polymer applies to the
grafting wall. In the following paragraph we explicitly compute
the pressure for ideal chains and relate it to the concentration
at the wall. We then argue that this also provides a good
pathway to compute the pressure applied by chains with excluded
volume.
\subsection{Gaussian grafted chain.}
\label{sec:2}
We consider a gaussian chain of N units, end-tethered by one
extremity to a non-adsorbing wall. The surface is described by
its height $ h (x,y)$, where $(x,y)$ denotes the position in the
horizontal coordinates frame. The thermodynamic
properties of the chain are described by the propagator
$G_{N}(\vec{r},\vec{r}\,')$, that satisfies the Edwards equation~\cite{doibook}
\begin{equation} 
\label{edwards} {\partial{G_{N}(\vec{r},\vec{r}\,')}\over
\partial{N}}={a^{2}\over 6}\Delta G_{N}(\vec{r},\vec{r}\,')
\end{equation}
with the following boundary conditions: $G_{N}(\vec{r},\vec{r}\,')\equiv 0$ on the
wall and $\lim_{N\rightarrow\ 0}
G_{N}(\vec{r},\vec{r}\,')=\delta(\vec{r}-\vec{r}\,')$. The length $a$ is
the monomer size. The statistical weigth of the chain attached
at a monomer distance from the origin $\vec{a}=(0,0,a)$ is given
by
\begin{equation}
\label{partition}
Z_N(\vec{a})=\int{d\vec{r}\,'\,G_{N}(\vec{a},\vec{r}\,')}
\end{equation}
the integral running over all the space available for the free end.
In the flat, reference case $h(x,y)=0$, the Green function can
be factorized
\begin{eqnarray}  
\label{green0} G_{N}^{(0)}(\vec{r},\vec{r}\,')  =
                      & ({3\over 2\pi
Na^{2}} )^{3/2} \exp\{-{3(x-x')^{2}\over 2Na^{2}}\}
\exp\{-{3(y-y')^{2}\over 2Na^{2}}\} \nonumber \\    &
\times (\exp\{-{3(z-z')^{2}\over
2Na^{2}}\}\!-\!\exp\{-{3(z+z')^{2}\over 2Na^{2}}\})  
\end{eqnarray}
and the partition function is
\begin{eqnarray}  
\label{part0}
Z_{N}^{(0)}(\vec{a}) & = & \int_{-\infty}^{+\infty}\!
dx'\int_{-\infty}^{+\infty}\! dy'\int_{0}^{+\infty}\!
dz'\,G_{N}^{(0)}(\vec{a},\vec{r}\,')   \nonumber \\
& = & \erf(\frac{a}{2R_{g}})
\end{eqnarray}
with $R_{g}=\sqrt{Na^{2}/6}$ the gyration radius of the chain,
and erf the error function~\cite{abrabook}. Now we seek for a
perturbative solution~\cite{marquesfournier} of the Edwards equation by performing a
small displacement $h$ of the surface. We write the partition
function as $Z_{N}=Z_{N}^{(0)}+Z_{N}^{(1)}+Z_{N}^{(2)}+\ldots$,
where
$Z_{N}^{(i)}$ is of order $h^{i}$ and $Z_{N}^{(0)}$ is defined
in equation~(\ref{part0}). One can notice that due to the linearity
of equation~(\ref{edwards}), each term of the perturbative expansion
obeys an Edwards equation
\begin{equation}  
\label{edpert}
{\partial{Z_{N}^{(i)}}\over\partial{N}}={a^{2}\over6}\Delta
Z_{N}^{(i)}
\mbox{ , } i=0,1,2,\ldots
\end{equation}
The solutions of successive orders are coupled through the
boundary conditions on the wall
\begin{eqnarray}  
\label{boundary} 0 & = & Z_{N}(x,y,h) \nonumber \\
  & = & Z_{N}(x,y,0) +h(x,y){ \partial{Z_{N}} \over
\partial{z}}(x,y,0) \nonumber \\ & & +\frac{h^{2}(x,y)}{2}{{\partial{^{2}Z_{N}}}\over
{\partial{z^{2}}}}(x,y,0)+\ldots
\end{eqnarray}
In the following, we concentrate on the first order term
$Z_{N}^{(1)} $, that is related, as we will see, to the pressure
exerted by the polymer on the surface. $Z_{N}^{(1)} $ is given
by the solution of equation~(\ref{edpert}) with the boundary condition
\begin{equation}
Z_{N}^{(1)}(x,y,0)=-h(x,y)\frac{{\partial{Z_{N}^{(0)}}}}
{\partial{z}}(x,y,0)
\end{equation}
The solution can then be written as~\cite{bartonbook}
\begin{equation}  
\label{magic} 
Z_{N}^{(1)}(\vec{a})  = 
\frac{a^{2}}{6}\int_{0}^{N} \!\!\!\!\!dn\!\!\int \! \!
dS'\,\frac{{\partial{G_{N-n}^{(0)}}}} {\partial{z'}}(x',y',0;\vec{a})
\,Z_{n}^{(1)}(x',y',0)
\end{equation}
so that the displacement of the surface is achieved to first order
in $h $ by the work
\begin{eqnarray}
\label{delf}
\Delta F & =&F[h]-F[0] \nonumber \\
         & =&-k_{B}T\log[1+\frac{Z_{N}^{(1)}}{Z_{N}^{(0)}}]
\nonumber \\
         & =&\int dSp(x,y)h(x,y)
\end{eqnarray}
where the function
$p(x,y)$ has the radially symetric form
\begin{equation}
\label{formpress}
p(r)=\frac{k_{B}T}{2\pi(r^{2}+a^{2})^{3/2}}(1+\frac{r^{2}+
a^{2}}{2R_{g}^{2}})\exp\{-\frac{r^{2}+a^{2}}{4R_{g}^{2}}\}   
\end{equation}
with $r=\sqrt{x^{2}+y^{2}}$. At point $\vec{r}=(x,y)$, the
elementary work $dF$ required to displace a volume
$dV(r)=h (r)dS$ is given by $dF=p(r)h (r) dS$. The function
$p(r)$ is therefore the pressure applied by the polymer on the
surface at point $r$. It is a non-homogeneous function - see figure~(\ref{pressplot})
- that sharply decays from its maximum value at the anchoring point
with the scaling form 
\begin{equation} 
\label{origin} p(r)\simeq\frac{k_{B}T}{2\pi r^{3}} \mbox{   for  }
a\ll r \ll R_{g}
\end{equation}
Well inside the polymer umbrella ($a\ll r \ll R_{g}$), where the
only relevant length is $r$, expression (\ref{origin}) is the
natural scaling for the pressure. In this region, most of the
monomers that contribute to the pressure are close to the
grafted end. For distances larger than the polymer size, the
pressure vanishes exponentially. A grafted polymer can then be
pictured as a microscopic pressure tool that applies a
well-defined but non-homogeneous force on a disk of radius $\sim
2R_{g}$ centered at the anchoring position. In the middle of the patch the pressure has a strong
value:
$ p(0) \simeq 2.4\times 10^{7}$ Pa for \(a=0.3 \)nm at room
temperature $T=25^{\circ}$C. A small area within a monomer
distance from the origin supports most of the total force $f$
exerted by the chain onto the surface
\begin{equation}
\label{force} f=\int_{0}^{\infty}2\pi rdrp(r)= {k_{B}T\over
a}\exp(-{a^{2}\over 4R_{g}^{2}}) \simeq 13.3
\mbox{pN}
\end{equation}
with the precedent values of monomer size and temperature. The grafted
monomer exerts a point like force $-f$  that ensures mechanical
equilibrium.   

Previous work~\cite{podgornik,lipowsky2} has focused on curvature contributions to the
polymer free energy $\Delta F$: by considering a surface of a
given shape, $\Delta F$ is calculated as a function of the
curvature ${1 \over R}$. For instance, for a sphere and a
cylinder one gets
$\Delta F_{sph}=-\sqrt{\pi}k_B T{R_{g}\over R} $ and $\Delta
F_{cyl}={1\over 2}\Delta F_{sph} $. The minus sign indicates
that the energy is lowered by spherical and cylindrical surfaces
that bend away from the polymer. We naturally recover these
results by considering a general surface defined by
$h(x,y)=-{x^{2}\over 2R_{1}}-{y^{2}\over 2R_{2}}$, and
evaluating the integral~(\ref{delf}) with $R_{1}=R_{2}=R$ for a
sphere and $R_{1}=R$ , $ R_{2}=0$ for a cylinder. Interestingly,
for a minimal surface
$(R_{1}=-R_{2}=R)$, there is no contribution of the curvature to
the free energy ($\Delta F =0$): to first order, the entropic
cost of tethering a gaussian chain to a plane or to a minimal
surface is the same. 

\subsection{Relation between pressure and concentration.}
\label{sec:3}
As explained above, the pressure that a grafted polymer exerts
on the wall has an entropic origin: by displacing the surface at
point $\vec{r} = (x,y)$ from its flat position $h
(\vec{r})$ one increases ($h <0$) or decreases ($h >0$) the
number of allowed chain configurations. The work per unit
surface associated with the corresponding entropy gain or loss
defines the pressure. Alternatively, the pressure can be viewed
as resulting from the forces applied by all the monomers at a
given surface point. If the surface potential acting on each
monomer is $u(z)$, the pressure is given by~\cite{eisen}
\begin{equation}
\label{prespot} p(x,y)=-\int_{0}^{\infty} dz {\partial{u} \over
\partial{z}}(z) c(x,y,z)
\end{equation}
with $c$ the monomer concentration. Expression (\ref{prespot})
reveals the linear relationship between pressure and
concentration at the wall but is not very usefull for the
continuous gaussian chain considered in this paper. Instead, we
directly derive the pressure-concentration relation by noting
that the monomer concentration of a chain tethered to a flat
surface is written as~\cite{degennesbook} 
\begin{equation}
\label{concentration} c^{(0)}(\vec{r}\,)={1\over
Z^{(0)}(\vec{a})}\int_{0}^{N} \!\!\! dn\int
d\vec{r}\,'G_{n}^{(0)}(\vec{a},\vec{r}\,)G_{N-n}^{(0)}(\vec{r},\vec{r}\,')   
\end{equation}
The interaction with the wall being purely repulsive,
one has $c^{(0)}(x,y,0)=0$ and
${\partial{c^{(0)}}\over \partial{z}}(x,y,0)=0$, the second
derivative of the concentration being the lowest derivative that
does not vanish on the surface. Rewriting equation (\ref{magic})
by taking into account the definition of the partition function
(\ref{part0}) leads to
\begin{equation}
\label{pressure} p(r)=k_{B}T{a^{2}\over
12}{\partial{^{2}c^{(0)}}\over
\partial{z^{2}}}(r,0)  
\end{equation}
with $r=\sqrt{x^{2}+y^{2}}$. Qualitatively, the pressure can be
associated with an ideal gaz pressure caused by the
concentration of monomers at a distance $z={a
\over \sqrt{6} }$ from the wall
\begin{equation}
\label{pc} p(r)=k_{B}Tc^{(0)}(r, z={a \over \sqrt{6} })
\end{equation}
Equivalently, equation (\ref{prespot}) can be used to assert
that the effective wall potential acting on the monomers has a
second moment of forces given by $\int_{0}^{\infty}
dzz^{2}{\partial{u} \over
\partial{z}}(z)=k_{B}T{a^{2} \over 6}$.  

\subsection{Grafted self-avoinding walks.}
\label{sec:4}
Flexible polymer chains in good solvent cannot be described by
gaussian statistics. They only exhibit ideal gaussian behaviour
close to the theta temperature, at which monomer attraction
compensates for steric repulsion. Above the theta point, the
polymers perform self-avoiding walks (SAW's) which lead to
distinct statistics. In particular the average dimension of a
SAW coil is larger than its gaussian equivalent, the end-to-end distance scales as
$R=N^\nu a$, with
$\nu$ the Flory exponent, close to $\nu\simeq 3/5$.

As stated in equation~(\ref{prespot}), the proportionality
between the polymer pressure and the monomer concentration in
the vicinty of the wall is expected to hold on general grounds,
independently of the approximations involved. Altough
equation~(\ref{prespot}) does not easily provide for a proportionality
coefficient, it does give a firm ground for predicting the
scaling form of the pressure applied to the wall by grafted
chains in a good solvent. For comparaison, we first recall the
structure of the monomer concentration profile for a gaussian
grafted polymer, a case  where an explicit calculation can be
performed~\cite{eisenbook}. The cone of equation
$z=r$ separates two regions in space. Outside the cone ($z\!
\ll \! r$) but well inside the polymer ``umbrella''($z,r\! \ll \! R_g$), the
monomer concentration grows quadratically from the wall, 
$c(r,z) \sim  z^2/ (r^3 a^2)  $. In this
region the wall has an important depletion effect on the polymer
configurations. Inside the cone ($z\!\gg \!r$) one recovers  the usual  bulk
concentration of a gaussian chain,
$c(r) \sim 1/(r a^2)$. The crossover between the two behaviours occurs
as one crosses the cone surface $z=r$. In the presence of excluded volume
interactions, the bulk concentration is given by   $c(r)
\sim 1/(r^{4/3} a^{5/3})$. For chains with excluded volume, the profile
grows from the wall as $c(r,z)\sim z^{5/3}$. Writing the scaling form in the region
$z\!\ll \!r$ that matches  bulk behaviour at $z=r$ leads to
$c(r,z) \sim z^{5/3}/( r^3 a^{5/3})$. The pressure applied by a
swollen grafted chain is therefore given by  $p(r)\sim k_{B}T
c(r,z=a) \sim k_{B}T r^{-3}$. It has the same scaling form as
the pressure applied by ideal chains. We expect such scaling to
be rather independent from the molecular details or from the differences between chain
models. 

Excluded volume effects might nevertheless  modify the amplitude and the range of the
applied forces. In order to quantify these effects we implemented a Monte-Carlo simulation on a
polymer attached to a flat, impenetrable wall. The chain is described as a pearl necklace
\cite{KKremer}, each pearl of size $a$. The first monomer is grafted to the wall with center of
mass position
$(0,0,0)$. During simulation a histogram $c(r,z)$ for the monomer center of mass concentration at
a distance $r$ along the wall and height $z$ above the wall is compiled. We are using a binsize
of $0.18\cdot a$ in the $r$ and $z$ direction.
To extract the concentration at the wall at a
distance $r$ away from the grafting point we fit the function
$f(z)$ = $c(r,z)$ with a fourth order polynomial in $z$ multiplied with an exponential
$\exp(-\lambda z)$, $\lambda$ a fitting parameter.
Upon extrapolating the fitted function to $z=0$
we get the concentration of monomers at the wall at a given distance $r$: $\lim_{z\rightarrow 0}
c(r,z)$. To ensure a reasonable error bar on the resulting concentration we
generated $6\cdot10^{6}$ configurations for gaussian random walks and self-avoiding random walks.
By analysing the statistics for end-to-end distance, which represents the slowest relaxing mode
for the chain, we estimate that we have an maximal error-bar of 12\% as $r$ is increased from a
few $a$, where the error-bar is less, to $30\cdot a$. 

We first consider
the Monte-Carlo results for 
a chain without excluded volume.  Because our Monte-Carlo chain
is actually a freely-hinged chain, we do not expect the
amplitude coefficient of the gaussian model to exactly hold. We
therefore plot in figure~(\ref{montefig}a),  both the expression for the
pressure from equation~(\ref{formpress}) and the values for the
monomer concentration at the wall extracted from
a Monte-Carlo simulation of a chain with $N=200$ monomers.  Agreement
is excellent, except at distances of the order of a monomer size where the
fixed length between monomers induces oscillations reminiscent of
those observed in the correlation function of hard spheres.  In
figure~(\ref{montefig}b), we show equivalent  results for  a chain of
N$=200$ monomers with excluded volume. Agreement is also excellent, if
we replace the dimension of the chain 
$R^2 = N a^2$ in expression~(\ref{formpress}) by $R^2 =
1.5  N^{2 \nu} a^2$, with $\nu = 0.6$. For
the chain representation  used in our simulations,  excluded volume
effects influence  only  the
range of the pressure field. The scaling form at small distances and even
the amplitudes are equivalent to those of the ideal chains.

\section{Polymers anchored on a  freely standing membrane.}
\label{sec:5}
Grafted polymers are small pressure patches: when a polymer is grafted
to a soft interface, the pressure deforms the interface into a
characteristic shape which is determined by the balance between the
pressure and the elastic response of the grafting surface. In this
chapter we consider first the deformation induced by a chain grafted on a
freely standing elastic membrane, and then membrane
induced interactions between two grafted chains.  

\subsection{Deformation induced by the pressure field of a gaussian
polymer.}
\label{sec:6}
The thermodynamic properties of a fluid membrane are well
described by the Canham-Helfrich Hamiltonian~\cite{helfrich1}, provided that the
thickness $d$ of the bilayer is small compared to the other relevant
lengths of the problem (i.e. $d \!\ll \!R_{g}$).  The surface is
described by its height $h(x,y)$, where $\vec{r}=(x,y)$ refers to the coordinate frame in the reference
plane $h(x,y)=0$.  Assuming a gentle surface deformation ($\mid\vec{\nabla}h \mid
\ll 1$), the Hamiltonian is written in the Monge representation as
\begin{equation}
\label{helfrich} H={\kappa\over 2}\int dxdy(\Delta h )^{2}
\end{equation}
with $\kappa$ the bending rigidity and $\Delta$ the
2-dimensional Laplacian operator. For a membrane with fixed
topology, the Gauss-Bonnet theorem implies that the gaussian
curvature term is constant and can therefore be ignored. For a
purely repulsive surface, the total free energy of the system
\{membrane+polymer\} is the sum of the bending energy of the
membrane and the work of the entropic force exerted by the
polymer
\begin{equation}
\label{polmem} F[h ]=F[0]+{\kappa\over 2}\int 2\pi rdr(\Delta
_{r}h )^{2}+\int 2\pi rdrp(r)h(r)
\end{equation}
with $F[0]$ the work required to graft the polymer to a flat plane.  Since
both the applied pressure and the boundary
conditions considered below are radially symmetric,  we coveniently
expressed all quantities in  cylindrical coordinates. The equilibrium shape
of the membrane ensues from a compromise between the applied pressure
and the restoring bending forces. Functional minimization of the free
energy with respect to the membrane profile
$h(r)$ leads to the Euler-Lagrange equation for the equilibrium shape
\begin{equation}
\label{eulagrange} 
\kappa
\Delta _{r}\Delta _{r}h (r)+p(r)=0
\end{equation}
where $\Delta _{r}={1 \over r}{d \over dr}r{d \over dr}$ is the
radial part of the Laplacian operator.  We first focus on the central region,
close to the grafting point, where most of the stress is concentrated. 
Here ($r\ll R_g$), the pressure behaves likes $k_{B}T/(2\pi r^{3})$, so that
the resulting deformation has a cone shape
\begin{equation}
\label{cone} 
h (r) \mathop{\simeq}_{r\to 0}-({k_{B}T\over 
\kappa}){r\over 2\pi}  
\end{equation}
independently of the boundary conditions. A point-like defect is
generated at the origin and we will refer to this conic shape as
 the {\em fundamental pinch} -- see figure~(\ref{pinchfig}).
  The surface
curvature diverges at short distances as $\Delta_r h
\mathop{\propto } r^{-1}$. Physically, this divergence is cut-off either
at a distance of the order of the membrane thickness $d$  by
 non-harmonic  terms in the curvature energy or, for infinitely thin
membranes, at the monomer length $a$ at which the pressure saturates.  
It is interesting to note that if a polymer is grafted to the tip of a purely
conic deformation with an arbritary slope, and then the slope determined by
balancing the {\sl global} entropy gain of the polymer and the elastic cost
of deforming the membrane into a cone shape, one finds the same slope
as that of expression~(\ref{cone}), given by the {\sl local} balance of
equation~(\ref{eulagrange})~\cite{lipowsky1}. This is an indication that the
conic region  supports most of the total stress imposed by the pressure
patch. 
Neglecting effects in the cut-off region, the analytical solution of
equation~(\ref{eulagrange}) can be written as
$ h (r)=h _{p}(r)+h _{bh}(r)
$, with $h _{bh}(r)=\frac{k_{B}T}{2\pi \kappa}[c_1+c_2\ln
(r)+c_3 r^{2}+c_4r^{2}\ln (r)]$ the kernel of the biharmonic operator
and 
\begin{eqnarray} h _{p}(r) = -{k_{B}T\over 2\pi \kappa}[{1\over
4}r\exp (-{r^{2}\over 4R_{g} ^{2}})-{\sqrt{\pi}\over
8}{r^{2}\over R_{g}} \erfc({r\over 2R_{g}})  \nonumber\\ +{\sqrt{\pi}\over
4}R_{g}\erf({r\over 2R_{g}})+
{R_{g}\sqrt{\pi }\over 2} \int_{0}^{{r\over R_{g}}}  {du \over u}
\erf({u\over 2}) ] 
\label{particuliere}    
\end{eqnarray}
a particuliar solution of equation~(\ref{eulagrange}). The constants
$c_1$, $c_2$, $c_3$ and $c_4$ are determined by the
boundary conditions. In the simple case considered here, the membrane
has no imposed constraints other than its known position of the center of
coordinates $h (0)= 0$. Because there are no forces acting  on the
membrane at large distances from the center the average curvature
must vanish there 
$\Delta_r h (r\to
\infty )=0$. This determines the four constants 
$c_1=c_2=c_3=c_4=0 $. At distances larger than
$R_{g}$, the profile is a catenoid, a radially symetric shape with zero
average curvature
\begin{equation}
\label{catenoid} h(r)\simeq -{k_{B}T \over 2\pi \kappa }R_{g}
\ln({r\over R_{g}})  \mbox{ for }  r\gg R_g
\end{equation}
The complete profile is displayed in figure~(\ref{pinchfig}). It has a
characteristic pinched shape, with a cone like deformation
(\ref{eulagrange}) that crosses over to the catenoidal shape
(\ref{catenoid}). The divergence of the profile is related to the
unconstrained nature of the membrane considered here. We will discuss in
section~\ref{sec:9} how the deformation profile is modified by the boundary conditions
or other external fields.

\subsection{Interaction potential between two grafted polymers.}
\label{sec:7}
Most often, bilayers anchor a finite concentration of
polymers. Each polymer is a pressure patch that carries with it a pinched form.  
Beyond the usual Van der Waals or steric  interactions between the
different chains, the superimposition of the different pinches will also
lead to membrane mediated forces.  Due to the linear nature of the pressure
contribution to the free-energy of the system, the many-body problem
reduces in this case to a sum of two body interactions that we now study.

The free energy of two pressure patches  applied at positions
$\vec{r}_1$ and $\vec{r}_2$, {\sl on the same side} of a membrane is
\begin{eqnarray} 
\label{twin}
F[h ,\vec r_{1},\vec r_{2} ]& = & \int dS[p(\mid\vec{r}-\vec{r}_{1}\mid )+p(\mid
\vec{r}-\vec{r}_{2}\mid )]h(\vec{r}) \nonumber \\ 
& &+{\kappa\over 2}\int dS(\Delta h(\vec{r}) )^{2} 
\end{eqnarray}
so that the deformation field obeys
\begin{equation} 
\label{edint}
\kappa \Delta \Delta h (\vec r)+p(\mid \vec r-\vec r_{1}\mid )+
p(\mid \vec r-\vec r_{2}\mid )=0
\end{equation}
with $\Delta=\Delta_{r}+{1\over r^{2}}{\partial^{2} \over
\partial
\theta^{2}}$ the Laplacian operator in cylindrical coordinates. Note however
that the problem has now lost its radial symmetry. A particular solution of this
equation is
$h_{p}(\mid
\vec r-\vec r_{1}\mid )+h_{p}(\mid
\vec r-\vec r_{2}\mid )$, the function $h_{p}$ being given by
(\ref{particuliere}). The general solution of the biharmonic
equation $ \Delta \Delta h =0$ that satisfies the requirement
$\lim_{r\rightarrow\ \infty}\Delta h =0 $ with a finite value at
the origin is simply a constant. If we impose the conditions $h
(\vec r_{1})=h (\vec r_{2})=0$, we are lead to the solution of
the differential equation (\ref{edint}) 
\begin{equation} 
\label{sol} h (\vec r)=h_{p}(\mid \vec r-\vec r_{1}\mid 
)+h_{p}(\mid \vec r-\vec r_{2}\mid ) - h_p(l) - h_p(0) \
\end{equation}
with $l$  the distance between
polymers, $l=\mid \vec{r_{1}}-\vec{r_{2}} \mid $. The interaction
potential
$V(l)$ is given by the difference between the free energy (\ref{twin}) and twice
the free energy of one isolated polymer. Inserting the solution of the Euler-Lagrange equation in 
(\ref{twin}) and integrating by parts leads to 
\begin{equation}
\label{vl}  
V_{curv}(l)=-\kappa \int dS\Delta h_{p}( \vec r )
\Delta h_{p}( \vec r-\vec l) \ 
\end{equation}
The  potential $V(l)$ is always negative: the interaction between two
polymers attached on the same side of the bilayer is attractive.
It also follows from equation~(\ref{vl}) that two
polymers anchored to {\sl the opposite side of a membrane} repel
each other. The interaction potential has in this case the same functional form 
but with the reverse sign. The interaction potential is shown in
figure~(\ref{potential}) for two polymers grafted on the same side.
Notice that the potential range is of the order of the polymer size: at larger distances, the
deformation field having a zero curvature shape,  the cost of grafting a second
polymer to it is the same as grafting a polymer to a flat interface. The
mechanisms responsible for attraction or repulsion are easy to understand.
When two polymers are grafted to the {\sl  same side} of a membrane, they can
both share the same deformation profile, instead of creating each one a profile
of their own. When they are grafted to {\sl opposite sides}, the deformations are
mutually neutralized: the polymers can only fully develop their deformation
profiles at a large distance from each other.

At short distances ($l\ll R_{g}$),  the attraction has 
a logarithmic behaviour
\begin{equation}
\label{inter} V_{curv}(l)\simeq  {(k_{B}T)^{2}\over 2\pi \kappa }\ln
(l/R_g) 
\end{equation}
For soft membranes the elastic constant is of order $k_{B}T$, which leads, for a
polymer with a radius of gyration of ten nanometers, $R_g=10$nm and a
minimum approaching distance of the order of a monomer size $l=0.3$nm, to an
attraction well of a couple of 
$k_{B}T$.  Moreover, the potential varies quadratically with temperature, this
class of interactions is thus quite sensitive to temperature variations. We will
further discuss in the conclusions the  possible implications of such sensitivity.

It is important to stress the differences between the potential that we
just described and other membrane induced  interactions abundantly
described recently~\cite{goulian,dommersnes}. The pressure patches do not lead to
Casimir-like interactions, there are here no logharitmic  or
algebraic tails. In the literature, the inclusions are considered as rigid bodies that
fixes the value of the membrane curvature at the inclusion site. It can be checked
that, if one consider very soft point-like inclusions for which the curvature
self-adjusts to a prefered value that minimizes the inclusion distortion and the
membrane bending, then there is no interaction potential between the
inclusions~\cite{fournierpc}. In the jargon of membrane induced interactions, our
potential corresponds  to a ``short range'' potential. Notice however that for
polymers, the range of the interaction can be at least one order of magnitude larger
($\sim$~tens of nanometers) than typical inclusion sizes ($\sim$~few nanometers).

The potential~(\ref{vl}) accounts only for curvature-mediated
interactions. At distances of order
$R_g$, inter-chain interactions give also rise to a repulsive contribution. We qualitatively
account for the polymer-polymer repulsion by separating the two chains with a mid-plane hard
wall~\cite{marquesfournier}. This over-estimated repulsive potential reads
$ -k_{B}T\ln[\erf(l/R_g)] $.  We now write the total interaction potential as
\begin{equation}
\label{interrep}  V(l)= V_{curv}(l) - b\, k_{B}T\ln[\erf(l/R_g)] \,
\end{equation}
with $b$ a constant smaller than unity. Both parts behave logarithmically at short distances: the
potential is attractive for small values of the bending rigidity and is repulsive for high values.
Figure~(\ref{potentialev}) shows the plots of $V(l)$ for different values of the rigidity $\kappa$ and
$b$ arbitrarily fixed at 
$b=1/(2\pi)$.
In this paticular case the crossover between attraction and repulsion occurs at
$\kappa=k_BT$. Changing the value of $b$ will accordingly rescale the crossover value. 
When the surface is covered with a finite density of chains, the onset of aggregation can be
monitored by the second virial coefficient $B$
\begin{equation}
\label{virial} B={1 \over 2}\int dS(1-\exp\{-{V(\vec{r}) \over k_BT}\})
\end{equation}
The plot of $B$ as a function of the bending rigidity with our particular choice of $b$ is shown
on figure~(\ref{virialfig}): for $\kappa < 0.6k_BT$, the second virial
coefficient becomes negative, indicating aggregation of different chains.

\subsection{Star-like polymer aggregates.}
\label{sec:8}

Attractive interactions between chains grafted to the surface may
lead to a star-like structure: the grafting points merge into a
core, while excluded volume repulsions acting on the arms give the
aggregate the hemispherical shape displayed in figure~(\ref{star}). We now
discuss how the structure of the aggregate changes the nature of the pressure
applied to the grafting surface, and the deformation that the pressure induces
on a free standing elastic membrane.

The structure of a star polymer with $f$ arms of polymerization index $N$ can be
described by the Daoud-Cotton model~\cite{daoud}. Attachment of the chains to a
central core effectively forces the local polymer density to be everywhere
inside the star above overlapping concentration. The star can therefore be
described as a semi-dilute solution~\cite{degennesbook}, with a local, position
dependent screening length
$\xi(r)$, where $r$ is here the distance from the center of the star in a frame of
spherical coordinates.  Pictorially, we represent this by associating with each
arm a string of blobs of  increasing size
$\xi(r)$. The radial dependence of the blob size $\xi(r)$ can be obtained   by
noticing that at a distance $r$ from the center there are  $f$ blobs of cross
section $\xi(r)^2$ occupying a total area of $4\pi r^2$. The blob size thus varies
as $\xi(r)\simeq r f^{-1/2}$ and the corresponding local polymer concentration
as $c_s(r) \simeq f^{2/3} a^{-5/3} r^{-4/3}$. Note that there is a crowded region
of size
$R_c\simeq a f^{1/2}$ in the middle of the star where the concentration reaches
one.  The size of the star can be obtained from monomer conservation  $N f
= 4
\pi\int_0^R r^2 \ dr c_s(r)$. Neglecting the small core region one gets
$R\simeq a N^{3/5} f^{1/5}$.

The structure of interest to us is a half-star, simply obtained  from the Daoud
and Cotton model by replacing $f/2$ arms by a repulsive half plane - see figure~(\ref{star}).
The local pressure is a simple function of the blob size $p_s(r)\sim k_B T \xi(r)^{-3}$,
leading to a pressure  applied {\sl on the surface} of the form
\begin{equation}
\label{presstar} 
p_s(r)\simeq f^{3/2} {k_B T \over r^{3}}\ ,
\end{equation}
if we now revert back to the previous
notation where $r$ is the distance from the center in a surface cylindrical frame. The reasons beyond this
functional form can also be related to the interfacial structure of the star. Close to the surface
there is a monomer depletion layer, growing from the surface as $c(r,z)\sim z^{5/3}$. Due to
screening, the bulk behaviour is recovered when one crosses the cone surface $z=rf^{-1/2}$.
Writting again the $r$ dependence of the depletion layer such as to match the bulk value
$c_s(r)\simeq f^{2/3}a^{-5/3}r^{-4/3}$, one gets
$c(r,z)\simeq p_s(r)(z/a)^{5/3}$, with $p_s(r)$ the pressure field in equation~(\ref{presstar}).
For distances
$r$ larger than the star size, we expect the pressure to vanish rapidly.

For ideal chains, the pressure applied by $f$ chains grafted at the
same point is $f$ times larger than the pressure applied by a single
chain. For polymers  with excluded volume there is an additional
crowding effect that results in a pressure $f^{3/2}$ times larger
than the pressure of a single chain. Also, the range of the pressure
 grows as the star size, and is a factor $f^{1/5}$ larger than the
range of a single chain.
Since the pressure field close to the grafting point has the same scaling form as
a single chain pressure, the patch still induces a conic deformation
on a free standing elastic membrane. The angle of the
cone is  more pronounced for a star than the angle of a pinch from a
single chain
\begin{equation}
\label{conestar} 
h_s (r) \simeq - f^{3/2} ({k_{B}T\over 
\kappa}) r  
\end{equation}
and the energy gained by the creation of the star-pinch is much greater than the
the sum of the energy gains of $f$ individual pinches
\begin{equation} {F_{curv}\over k_{B}T} \sim -{k_{B}T\over
\kappa }f^{3}\ln({R\over a})
\end{equation}
The Daoud-Cotton model also allows to compute the excluded volume cost to
build a star-like structure, the result is 
\begin{equation} {F_{ev}\over k_{B}T} \sim
f^{3/2}\ln({R\over a})
\end{equation}
For a number of chains $f$ greater than a threshold $f_0\sim (\kappa
/k_{B}T)^{2/3}$, aggregation is always favored. Nevertheless, that process might be slow since it is 
hindered by an energy barrier $\Delta F\sim \kappa$. It is also important to stress that the
mecanical constraints on the bilayer may limit the maximum aggregation number. This can be
determined by setting
$\mid \vec{\nabla} h \mid \sim 1$ in expression~(\ref{conestar}) which leads to $f_{\max}\sim
f_0$. At this stage a piece of membrane decorated with $f_{\max}$ polymers might as well detach
from the  main membrane, leading to a coexistance between decorated membranes and decorated small
vesicles or micelles~\cite{joannic}.

\section{Polymers anchored on a constrained membrane.}
\label{sec:9}
In most practical situations, the bilayer does not stand free in the solvent but it
is  subjected to additional constraints. In this section, we consider two
important situations. First we discuss the pressure applied by a grafted
polymer to a supported bilayer. Supported bilayers are fluid membranes  that
adhere to a substrate, allowing for instance for the observation of
cell phenomena like membrane protein aggregation, opening of ionic
channels and others~\cite{seifert,albersdorfer,bernard}.  Adhesion of membranes on a substrate  plays a 
role in biological phenomena such as endocytosis and exocytosis~\cite{albertsbook}, it is also
of relevance in biotechnological processes, such as drug delivery by
liposomes~\cite{lasic}. A second important case of constraint membranes
corresponds to lamellar $L_{\alpha }$ phases, where the membranes  are confined
by interactions with their neighbors in the lamella stack~\cite{warriner,castro}. 

\subsection{Supported bilayers.}
\label{sec:10}
We consider a membrane adhering to a
flat surface with a contact energy per unit area $\Gamma /2$.
Application of a pressure patch peels off a region of radius $L$ and
central height $h_{0}=h (r=0)$. The corresponding free energy
functional is
\begin{eqnarray}
\label{enersb} 
F[h(r),L,h_{0} ]= F_{0}-\Gamma \pi
L^{2}+{\kappa\over 2}\int_{0}^{L} \! 2\pi rdr(\Delta _{r}h
)^{2}  \nonumber \\ +\int_{0}^{L} \! 2\pi rdrp(r)(h(r) -h_{0} ) 
\end{eqnarray}
with $F_{0}$ a constant. Notice that the membrane height $h$ is here
measured with respect to the substrate. Performing the functional minimization
of equation~(\ref{enersb}) with respect to the deformation 
$h$ leads to an Euler-Lagrange differential equation identical to
equation~(\ref{eulagrange}) of the free-standing case, with boundary conditions
$h (0)=h_{0}$, $h (L)=0 $ and $ \frac{dh }{dr}(L)=0 $. Further minimization of
the free energy~(\ref{enersb}) with respect to the size of the peeled region $L$, and
with respect to the heigth at the origin $h_0$,  provides two more boundary
conditions 
\begin{eqnarray}
\label{lapl}
\frac{\partial F}{\partial L}=0 &\Leftrightarrow &\Delta h
(r=L)=(\frac{\Gamma }{\kappa } )^{1/2}\\
\label{dlapl}
\frac{\partial F}{\partial h_{0}}=0 &\Leftrightarrow
&\frac{d}{dr}\Delta h (r=L)=0
\end{eqnarray}
Relation (\ref{lapl}) equates the balance between the
attractive potential and the elastic moment at the contact line
$r=L$~\cite{landau7}. 
The peeled radius is given by the implicit equation for $L$
\begin{equation}
\frac{1}{4}\exp(-{L^{2}\over 4R_{g}^{2}})-{\sqrt{\pi }\over
2}{R_{g}\over L}\erf({L\over 2R_{g} })+\pi {L\over R_{g}}\beta=0
\end{equation}
The parameter $\beta^2 =\kappa \Gamma R_{g}^{2} /(k_{B}T)^{2}$ controls the
value of $L$. For small adhesive energies or very flexible membranes we
have $\beta
\ll 1$, the membrane is loosely attached to the surface and the
deformation is similar to the free-standing case:
$L\!=\!R_g /(2\sqrt{\pi} 
\beta )^{1/2}$ and $h_{0}\!=\!(k_BT/4\sqrt{\pi} \kappa )R_g\ln (L/R_{g})$.
For large adhesive energies or stiff membranes, $\beta
\gg 1$, only the conical deformation survives and the pinch
height is proportional to the peeled radius:  $L=R_g /(2\pi \beta)$   and $h_{0}=(3k_BT/
16
\pi\kappa)L$. One migth
be astonished that these quantities cannot be expressed  only as a function of  the
lenght
$\lambda_0 =(\kappa/\Gamma)^{1/2}$. For many problems involving bending energies and
adhesion (or interfacial tension) of membranes this length separates two regimes.
On length scales larger than $\lambda_0$,  adhesion or tension effects dominate the
 behaviour of the membrane, while for lenghts below $\lambda_0$ the deformation is
ruled by the bending curvature.  In our case, the results can also be understood in
terms of $\lambda_0$,  by recalling first the simpler case of a membrane that adheres
to the surface but has a fixed, given heigth
$\zeta_0$ at the origin. It is easy to show that in that case the peeling radius is
given by $L = (8 \lambda_0^2 \zeta_0^2)^{1/4}\sim (\lambda_0 \zeta_0)^{1/2}$. But
when a polymer patch is applied, the heigth is fixed by the polymer pressure and it
follows, for strong adhesions, the conic form~(\ref{cone}). One has thus $\zeta_0
\sim L/\kappa $ which leads to $L\sim \lambda/\kappa $. This holds up to a length $L$ of the
order of the radius of gyration $R_g$. For smaller adhesion strengths the balance
is determined by $\zeta_0 \sim R_g/\kappa $ leading to $L\sim (R_g \lambda_0 /\kappa )^{1/2} $.

Adhesion energies can be found in the range $\Gamma \sim
10^{-7}-10^{-4}$ mN.m$^{-1}$. For typical bending modulii $\kappa \sim
5-20k_{B}T$, the extension of the peeled zones is in the range of
one to ten nanometers, which is also the typical size for polymers. 

\subsection{Membrane in a potential well.}
\label{sec:11}

In this section, we focus on the effect of pressure patches applied to
membranes confined in an harmonic potential. This is for instance relevant to
describe  the lyotropic smectic phases $L_\alpha $~\cite{safranbook} but serves also as a paradigm
for other situations where the membrane is constraint by an external soft
potential.  The energy functional of one bilayer reads
\begin{eqnarray}
\label{smect}
 F[h(r) ]=  F[0]+\int_{0}^{\infty } 2\pi
rdrp(r)(h(r) -h_{0} )  \nonumber\\  +{\kappa\over
2}\int_{0}^{\infty } \! 2\pi rdr(\Delta_{r}h  )^{2} +{B\over
2}\int_{0}^{\infty } \! 2\pi rdrh (r)^{2}
\end{eqnarray}
where $h_{0}$ is the value of the deformation at the origin. In the case of a
stack of membranes the amplitude of the harmonic potential $B$ is  the 
compression modulus of the system and can be related to the curvature of the
interaction potential. The natural length arising in
expression~(\ref{smect})  is
$l_{0}=({\kappa \over B})^{1/4}$. On scales larger than $l_0$  the deformation 
is controled by the harmonic potential while on smaller scales the behaviour of
the membrane is ruled by curvature effects. In particular, for non-ionic $L_\alpha$
phases where steric Helfrich interactions control the repulsion between the
membranes~\cite{helfrich2,safinya}, it is possible to show that the lenght
$l_0$ is proportional to the average interlamellar spacing $d$
\begin{equation} l_{0}=({\pi\over 6})^{1/2}({\kappa \over
k_{B}T})^{1/2}d
\end{equation}
Minimizing the free energy functional with respect to $\zeta = h-h_0$
leads to
\begin{equation}
\label{harm}
\kappa \Delta_{r}\Delta_{r}\zeta (r) +B\zeta (r) =-(p(r)+Bh_{0})
\end{equation}
with boundary conditions are $\zeta (0) =0
$ and $\zeta (\infty )=-h_0 $. In order to solve equation~(\ref{harm}),
we first evaluate the associated Green function $g(r,r')$
defined by
\begin{equation}
\kappa \Delta_{r}\Delta_{r}g(r,r')+Bg(r,r')=\delta (r-r')
\end{equation}
so that the deformation field $\zeta$ is eventually obtained from
\begin{equation} \zeta (r)=-\int_{0}^{\infty }
dr'g(r,r')(p(r)+Bh_{0})
\end{equation}
A straightforward but tedius calculation gives for $r>r'$
\begin{eqnarray} g_{+}(r,r')& = & -{k_{B}T\over
\kappa}l_{0}^{2}r'\{\Bei_{0}(r'/l_{0})\Ker_{0}(r/l_{0})
\nonumber \\
 & & +\Ber_{0}(r'/l_{0})\Kei_{0}(r/l_{0})  \\ & & +{4\over
\pi}\Kei_{0}(r'/l_{0})\Kei_{0}(r/l_{0})\} \nonumber
\end{eqnarray}
and for $r<r'$
\begin{eqnarray} g_{-}(r,r')& = & -{k_{B}T\over
\kappa}l_{0}^{2}r'\{\Bei_{0}(r/l_{0})\Ker_{0}(r'/l_{0})
\nonumber \\ & & +\Ber_{0}(r/l_{0})\Kei_{0}(r'/l_{0})  \\
 & &  +{4\over \pi}\Kei_{0}(r/l_{0})\Kei_{0}(r'/l_{0})\} \nonumber
\end{eqnarray}
where the Kelvin functions $\Ber_{0}(x)$, $\Bei_{0}(x)$,
$\Ker_{0}(x)$ and $\Kei_{0}(x)$ are the real and imaginary parts of
the modified Bessel functions
$ \mbox{I}_{0}(xe^{i\pi/4}) $ and $ \mbox{K}_{0}(xe^{i\pi/4})~\cite{abrabook} $. The harmonic potential
allows then for an oscillatory profile, as depicted on figure~(\ref{harmonic}).

The amplitude of the deformation, $h_0$ is obtained by minimizing the free energy with respect to
$h_0$, leading to
\begin{equation}
h_{0}=l_{0}{k_{B}T\over
8\kappa}\int_{0}^{\infty}{dx\over
x^2}(1+\frac{4}{\pi}\Kei_{0}(x))\exp(-\frac{x^2}{4\alpha^2})(1+{x^2\over
2\alpha^2})
\end{equation} 
where $\alpha=R{g}/l_{0}$ . Asymptotically $h_0$ varies like the logarithm of $\alpha$ for small
values of $\alpha$, and decays with a power law for large values: $h_0\sim \alpha^{-2}$ - see
figure~(\ref{h0}).
For Helfrich systems where the length
$l_0$ is proportional to the interlamellar spacing $l_0\sim d$ (and $\kappa \sim k_{B} T$), one
may easily induce a deformation of
$0.1d$, by using polymers of gyration radius $R_g\sim 0.5 d$.  For even smaller
polymers the amplitude of the deformation becomes comparable to the polymer
size. For polymers much larger than the interlamellar distances, our approach
would need to be completed by accounting also for the pressure exerted by the polymer on the
neighboring membranes (of order of
$k_{B}T d^{-3})$~\cite{bickel1}. 

\section{Discussion and conclusion.}
\label{sec:12}
We have developed in this paper a new picture to describe the interactions
between a grafted polymer and a flexible membrane. 

We have shown that the
polymer behaves in fact as a {\sl small pressure tool}. The pressure applied by
this tool is very high close to the grafting point and decays sharply over a
distance of order
$R_{g}$, the gyration radius of the chain. Through scaling arguments and
numerical simulations we further confirmed that
excluded volume interactions mainly change the range of the pressure field, while its
amplitude and functional form are rather independent of solvent conditions. 

Under the pressure of the polymer, a flexible interface assumes a characteristic
deformation: each flexible surface is  a pressure sensor. We have shown that
for fluid membranes, the deformation has a pinched  conic form. The exact shape
of the pinch depends on the boundary conditions imposed upon the membrane, but
its form at the center of the deformation field is rather universal.

For many grafted polymers, the deformation field induces an interaction potential
between the polymers. We have shown that this potential is attractive for
polymers grafted on the same side of a membrane and repulsive for polymers
grafted on opposite sides. This interaction potential rises interesting
possibilities. For instance, by changing the temperature one might expect to
control the aggregation behavior of polymers grafted on the membrane. Also, if
many polymers are added to a vesicle, the attraction might bring many polymers
to the same site, increasingly catastrophically the pressure until a decorated
micelle or small vesicule detaches from the surface.

Our treatment of the polymer induced
deformations does not account for the fluctuations of the membrane.  Thermal
 fluctuations arise spontaneously in membrane systems on the scales above
the  de~Gennes-Taupin persistence length: $\xi=a'\exp({4\pi \kappa\over k_{B}T})$~\cite{degennes}.
When the elastic constants are of the order of $k_{B}T$, the lenght $\xi$ is in the
range $10-100$nm, and fluctuations become important. At the level of the first
order perturbation scheme implemented in this paper, the pressure is an external field that only couple
linearly to the deformation of the membrane, so the fluctuation spectrum is not perturbed. It is therefore
important to extend to second order this type of perturbative expansion in order to determine both the
corrections to the elastic parameters of the membrane and the corrections to the fluctuation spectrum
itself.


{\bf Acknowledgments:} We gratefully aknowledge inspiring
discussions with L. Auvray, J.-B. Fournier, S. Obukov and T.A. Witten.



\newpage

{\bf FIGURES}

\vskip 1truecm

\begin{figure} [t]
\caption{The pressure applied by a grafted polymer to the
surface. We chosed here $a=0.1 R_g$ and arbitrary units of
pressure. The insert stresses the scaling form $ r^{-3}$ close to the grafting point.} 
\label{pressplot}
\end{figure}

\begin{figure} [t]
\caption{Comparaison of the calculated pressure and monomer concentration at 
the wall extracted from Monte-Carlos simulations. a) Freely-hinged chain of 200
monomers. b) Freely-hinged chain of 200 monomers with excluded volume. The
continuous line is the expression~(\ref{pressure}) of the pressure. Note that there is no
adjustable parameter for the gaussian chain. For the chain with excuded volume, the size of the
polymer is $R^2=1.5N^{2\nu}a^2$, with $\nu=0.6$.} 
\label{montefig}
\end{figure}

\begin{figure} [t]
\caption{The pinched form of a fluid membrane under the
pressure of a grafted polymer. The deformation profile is the
actual calculated form  for a freely standing membrane.} 
\label{pinchfig}
\end{figure}

\begin{figure} [t]
\caption{The membrane-mediated interaction potential between two
gaussian polymers, grafted to the same side of a membrane. Polymers grafted on opposite sides
have the same functional form with the reverse sign.} 
\label{potential}
\end{figure}

\begin{figure} [t]
\caption{The two-body interaction potential with the estimated inter-chain repulsion. The
plots show ${2\pi V(l)\over k_BT}$ as a function of $l/R_g$ for different values of the bending
rigidity (in $k_BT$ units), with fixed parameter $b=1/(2\pi)$ .} 
\label{potentialev}
\end{figure}

\begin{figure} [t]
\caption{The second virial coefficient as a function of the bending rigidity (in $k_BT$ units),
with fixed paramater $b=1/(2\pi)$. Chains aggregation is induced by negative $B$, which occurs
at
$\kappa < 0.6k_BT$ for this particular value of $b$.} 
\label{virialfig}
\end{figure}

\begin{figure} [t]
\caption{The conformation of a star anchored to a plane. Inside each ``blob'' of size $\xi
(r) \sim rf^{-1/2}$ the branches have a single-chain behaviour.}
\label{star}
\end{figure}

\begin{figure} [t]
\caption{The deformation profile of a fluid membrane for different strengths of the
harmonic potential. The curves are plotted for $\alpha=0.5$, 1 and 2, the largest
deformation corresponding to the weaker potential.} 
\label{harmonic}
\end{figure}

\begin{figure} [t]
\caption{Amplitude of the pinch at the grafting point. The parameter $\alpha$ controls the
strength of the potential. For a soft potential, $h_0$ varies like the logarithm of $\alpha$.
The insert emphasizes the scaling form for large values of $\alpha$.} 
\label{h0}
\end{figure}


\begin{thebibliography}{1}

\bibitem{israelbook}
J.N. Israelachvili, {\em Intermolecular and surface forces} (Academic Press, New York), 1992.

\bibitem{albertsbook}
B. Alberts, D. Bray, A. Johnson, J. Lewis, M. Raff, K. Roberts and P. Walter, {\em Molecular Biology of the
Cell} (Garland Publishing, New York), 1998.

\bibitem{vandepas}
J.C. van de Pas, {\em Colloids and Surfaces A} {\bf 85} (1994), 221.

\bibitem{warriner}
H.E. Warriner, S.H.J. Idziak, N.L. Slack, P. Davidson and C.R. Safinya , {\em Science} {\bf 271} (1996),
969.

\bibitem{yiyang}
Y. Yang, R. Prudhomme, K.M. McGrath, P. Richetti and C.M. Marques, {\em Phys. Rev. Lett.} {\bf 80} (1998),
2729.

\bibitem{bouglet}
G. Bouglet, C. Ligoure, A.-M. Bellocq, E. Dufourc and G. Mosser, {\em Phys. Rev. E} {\bf 57}
(1998), 834.

\bibitem{joannic}
R. Joannic, L. Auvray and D.D. Lasic, {\em Phys. Rev. Lett.} {\bf 78} (1997),  3402.

\bibitem{ringsdorf}
H. Ringsdorf, J. Venzmer and F. Winnik, {\em Angew. Chem. Int. Ed. Engl.} {\bf 30} (1991),  315.

\bibitem{cates}
M.E. Cates, {\em Nature} {\bf 351} (1991),  102.

\bibitem{frette}
V. Frette, I. Tsafir, M.-A. Guedeau-Boudeville, L. Jullien, D. Kandel and J. Stavans, {\em Phys. Rev. Lett.}
{\bf 83} (1999),  2465.

\bibitem{helfrich1}
P.B. Canham, {\em J. Theoret. Biol.} {\bf 26} (1970), 61. W. Helfrich, {\em Z. Naturforsch.} {\bf 28c}
(1973),  693.

\bibitem{safranbook}
S. Safran, {\em Statistical Thermodynamics of Surfaces, Interfaces and
  Membranes} (Addison-Wesley, Reading, MA), 1994.

\bibitem{cantor}
R. Cantor, {\em Macromolecules} {\bf 14} (1981),  1186.

\bibitem{milner}
S.T. Milner, T.A. Witten and M.E. Cates {\em Macromolecules} {\bf 22} (1989),  853.

\bibitem{podgornik}
R. Podgornik, {\em Europhys. Lett.} {\bf 21} (1993),  245.

\bibitem{lipowsky1}
R. Lipowsky, {\em Europhys. Lett.} {\bf 30} (1995),  197.

\bibitem{hanke}
A. Hanke, E. Eisenriegler and S. Dietrich, {\em Phys. Rev. E} {\bf 59} (1999),  6853.

\bibitem{bickel2}
T. Bickel, C. Jeppesen and C.M. Marques, to appear in {\em Phys. Rev. E }.

\bibitem{breidenich}
During the preparation of this manuscript, an independent work by M. Breidenich, R.R. Netz and R.
Lipowsky appeared in {\em Europhys. Lett.} {\bf 49} (2000), 431, describing some of the results
presented in our paper.

\bibitem{degennesbook}
P.-G. deGennes, {\em Scaling Concepts in Polymers Physics} (Cornell University
  Press, Ithaca, NY), 1979.

\bibitem{eisenbook}
E. Eisenriegler, {\em Polymers near surfaces} (World Scientific, Singapore), 1993.

\bibitem{doibook}
M. Doi and S.F. Edwards  {\em The theory of polymer dynamics} (Clarendon
  Press, Oxford), 1986.

\bibitem{abrabook}
M. Abramowitz and I.A. Stegun  {\em Handbook of Mathematical Functions} (National Bureau of Standards,
Washington, DC), 1964.

\bibitem{marquesfournier}
C.M. Marques and J.-B. Fournier, {\em Europhys. Lett.} {\bf 35} (1996),  361.

\bibitem{bartonbook}
G. Barton, {\em Elements of Green's functions and propagation} (Clarendon
  Press, Oxford), 1989.

\bibitem{lipowsky2}
C. Hiergeist and R. Lipowsky, {\em J. Phys. II France} {\bf 6} (1996),  1465.

\bibitem{eisen}
E. Eisenriegler, {\em Phys. Rev. E} {\bf 55} (1997),  3116.

\bibitem{KKremer} K. Kremer, in {\em Computer Simulations in Chemical Physics}, M.P. Allen and
D.J. Tildesley eds. (Kluver Academic Publishers), 1993, pg. 397-459 

\bibitem{goulian}
M. Goulian, R. Bruinsma and P. Pincus, {\em Europhys. Lett.} {\bf 22} (1993),  145.

\bibitem{dommersnes}
P.G. Dommersnes, J.-B. Fournier and P. Galatola, {\em Europhys. Lett.} {\bf 42} (1998),  233.

\bibitem{fournierpc}
J.-B. Fournier (private communication).

\bibitem{daoud}
M. Daoud and J. Cotton, {\em J. Phys. (Paris)} {\bf 43} (1982),  531.

\bibitem{seifert}
U. Seifert and R. Lipowsky, {\em Phys. Rev. A} {\bf 42} (1990),  4768.

\bibitem{albersdorfer}
A. Albersd\"orfer, R. Bruinsma and E. Sackmann, {\em Europhys. Lett.} {\bf 42} (1998),  227.

\bibitem{bernard}
A.-L. Bernard, M.-A. Guedeau-Boudeville, L. Jullien and J.-M. DiMeglio, {\em Europhys. Lett.} {\bf 46}
(1999),  101.

\bibitem{lasic}
D.D. Lasic, {\em Tibtech} {\bf 16} (1998),  307.

\bibitem{castro}
F. Castro-Roman, G. Porte and C. Ligoure, {\em Phys. Rev. Lett.} {\bf 82} (1999),  109.

\bibitem{landau7}
L.D. Landau and E.M. Lifshitz, {\em Theory of Elasticity} (Butterworth-Heinemann, Oxford), 1998.

\bibitem{helfrich2}
W. Helfrich, {\em Z. Naturforsch.} {\bf 33a} (1978),  305.

\bibitem{safinya}
C.R. Safinya, D. Roux, G.S. Smith, S.K. Sinha, P. Dimon, N.A. Clarck and A.-M. Bellocq, {\em Phys. Rev.
Lett.} {\bf 57} (1986),  2718.

\bibitem{bickel1}
T. Bickel, C. Jeppesen and C.M. Marques, to appear in {\em C. R. Acad. Sci. Paris }.

\bibitem{degennes}
P.-G. deGennes and C. Taupin, {\em J. Phys. Chem.} {\bf 88} (1982),  2294.


\end{thebibliography}
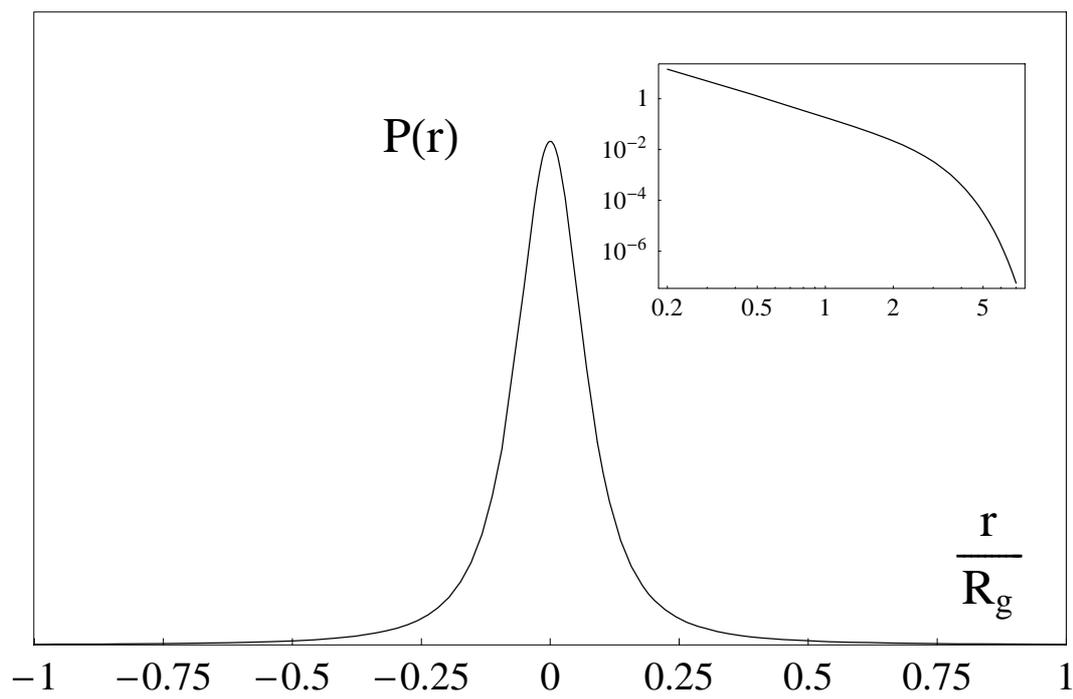
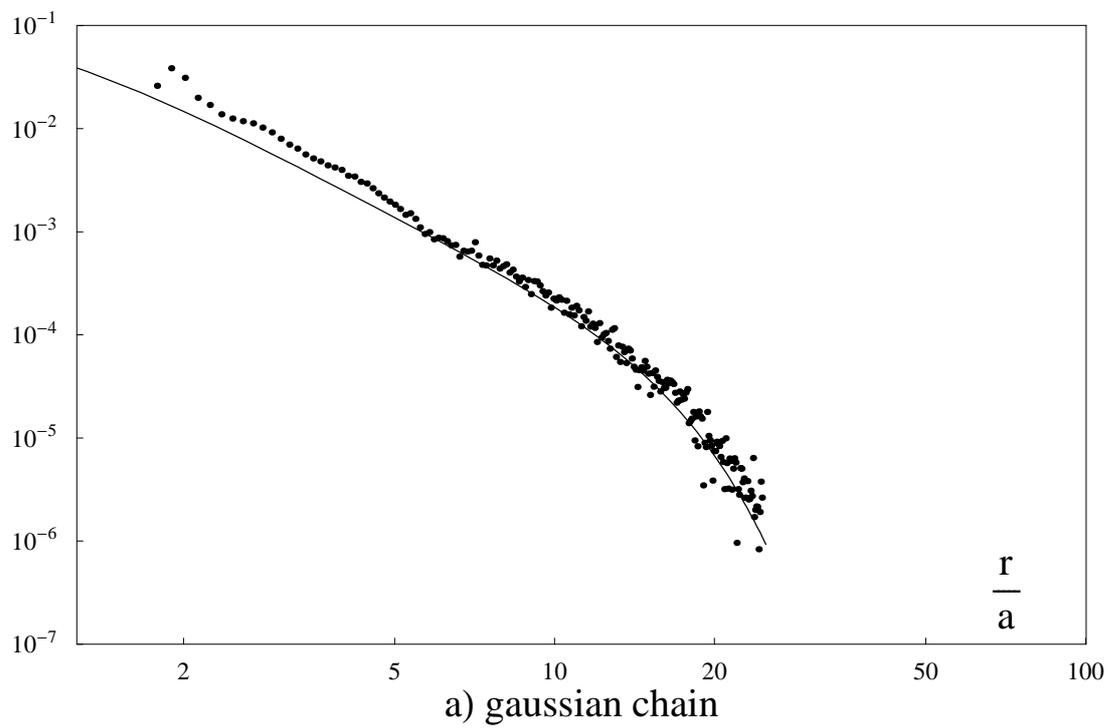
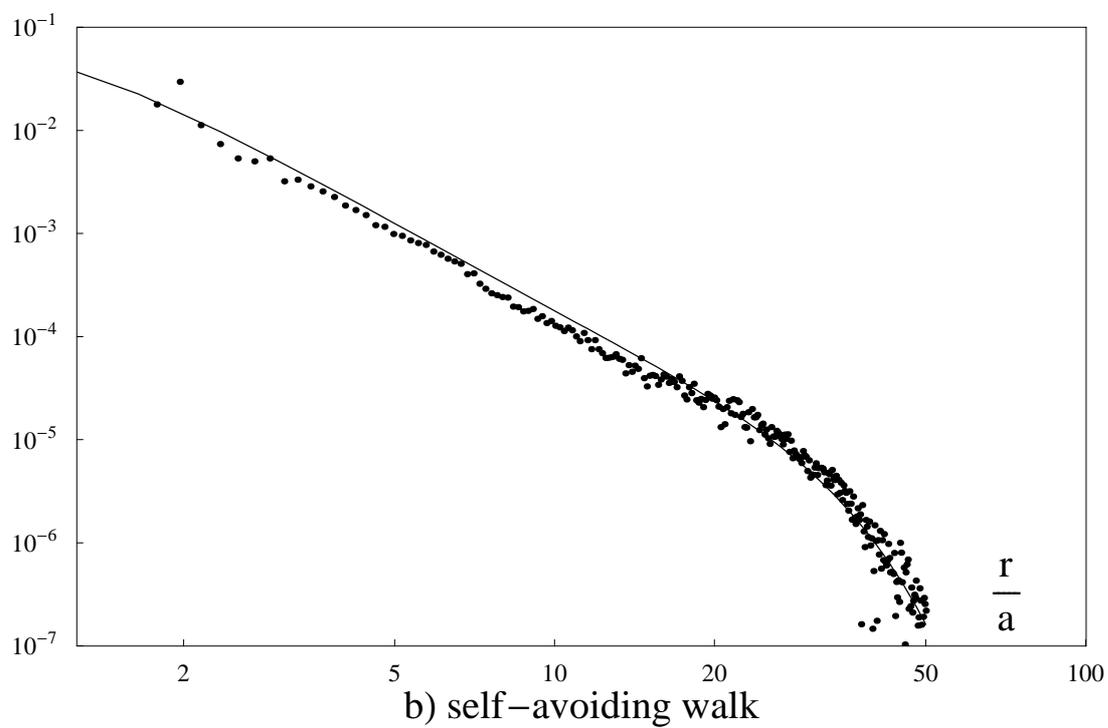
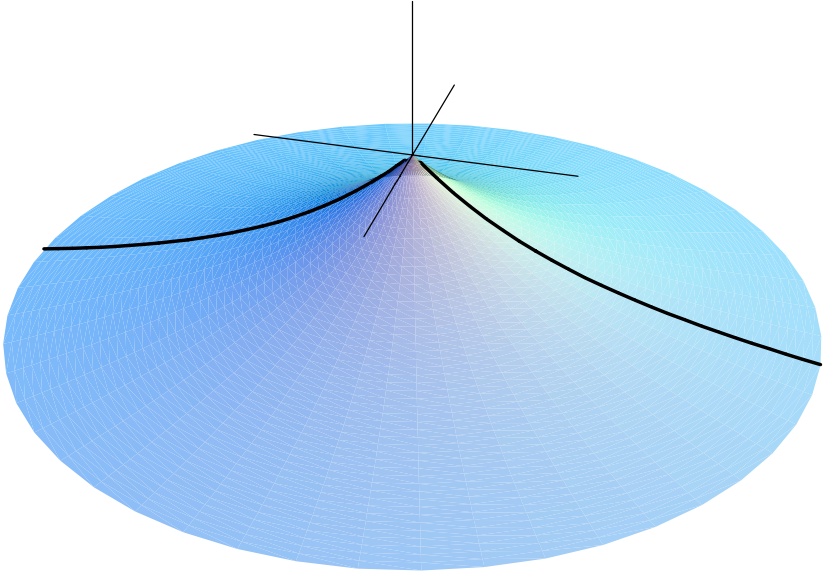
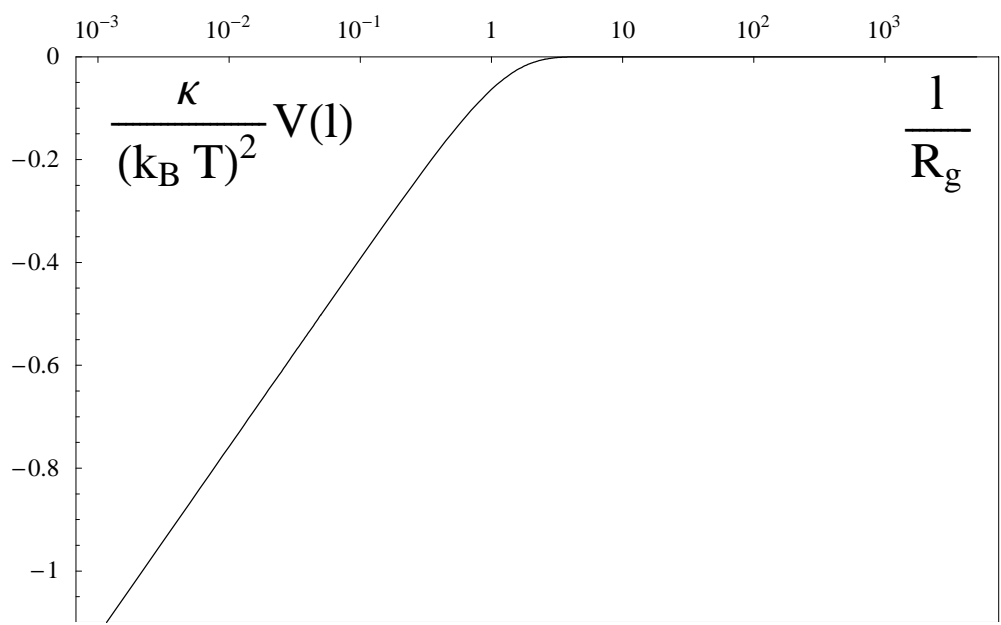
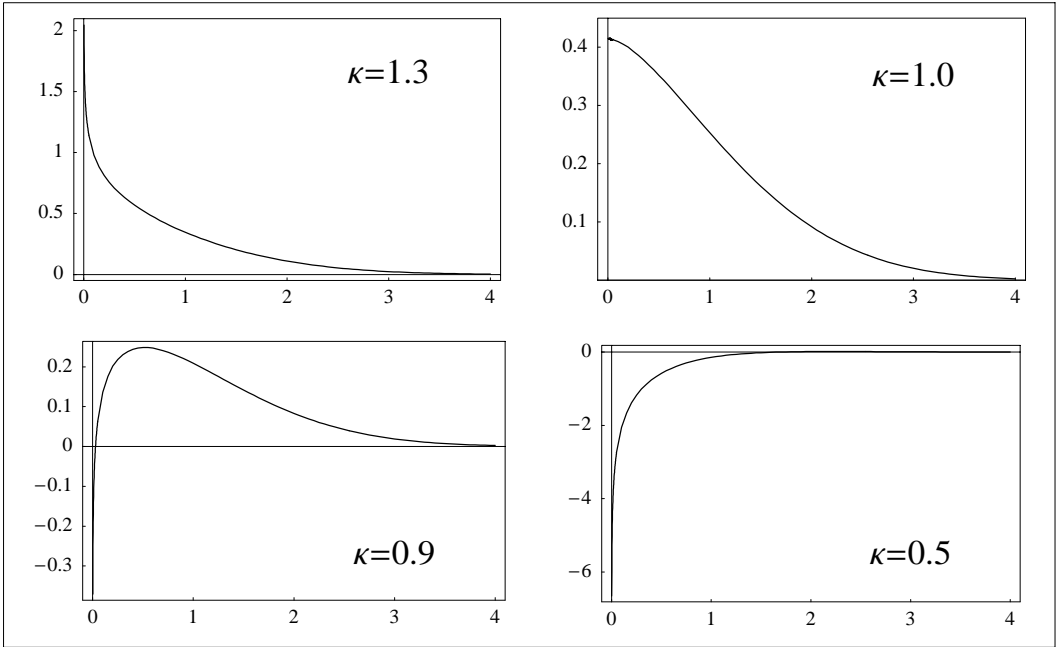
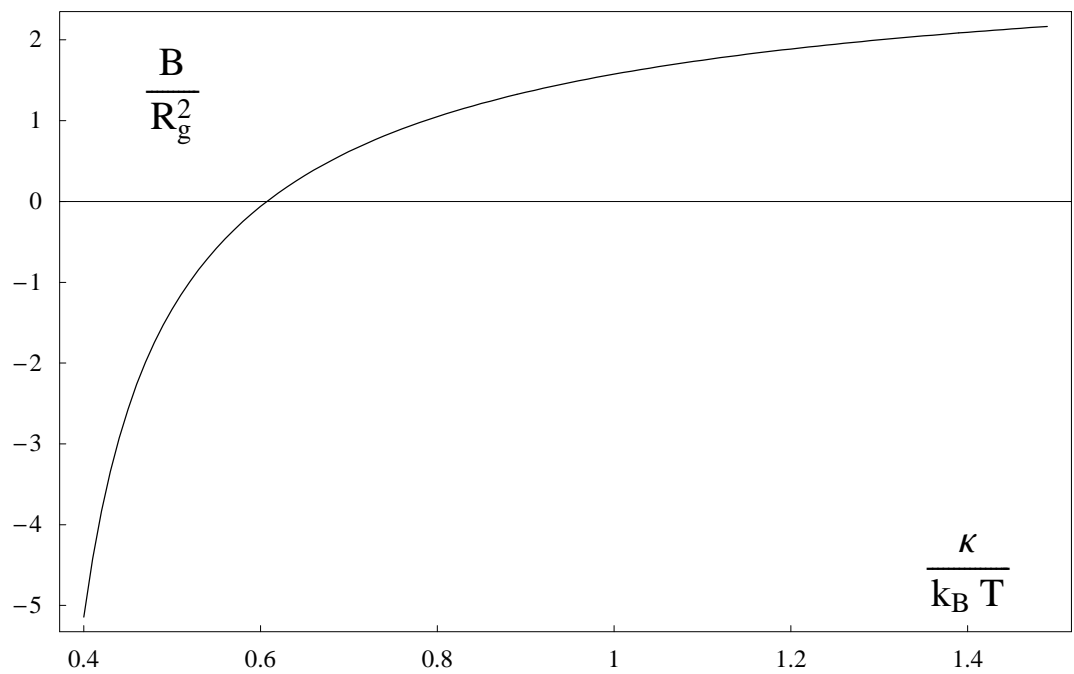
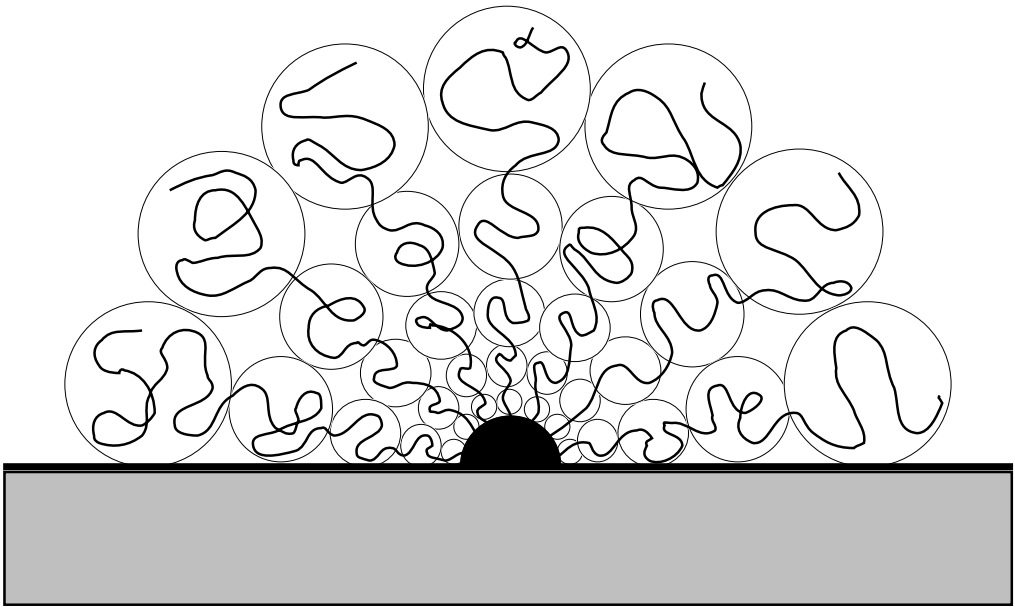
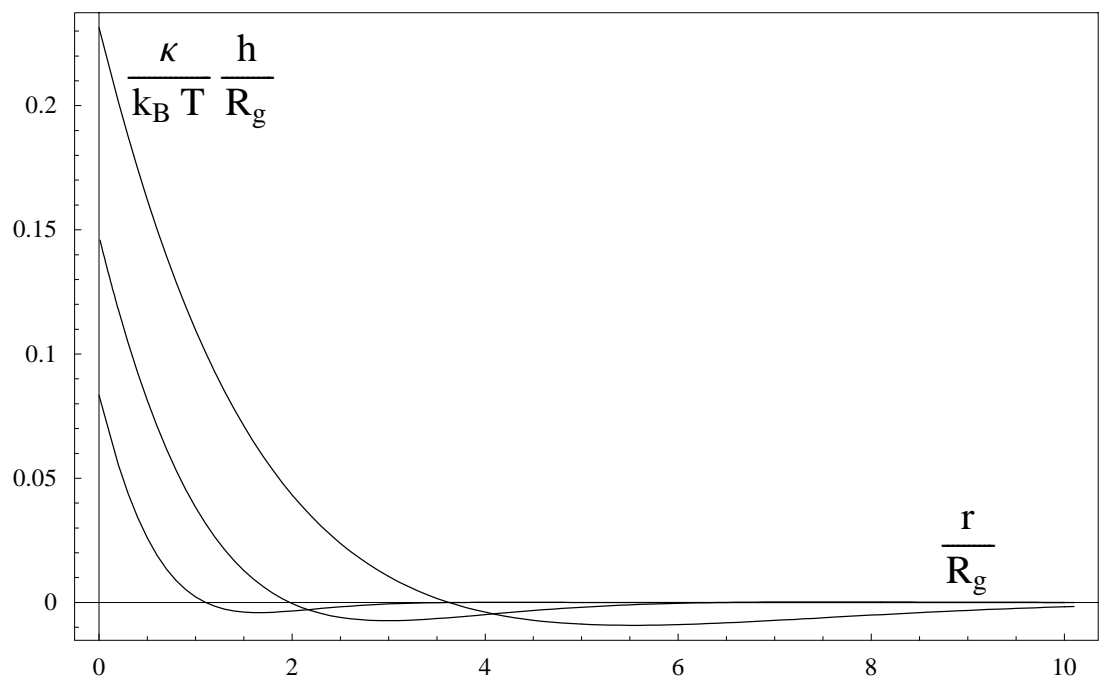
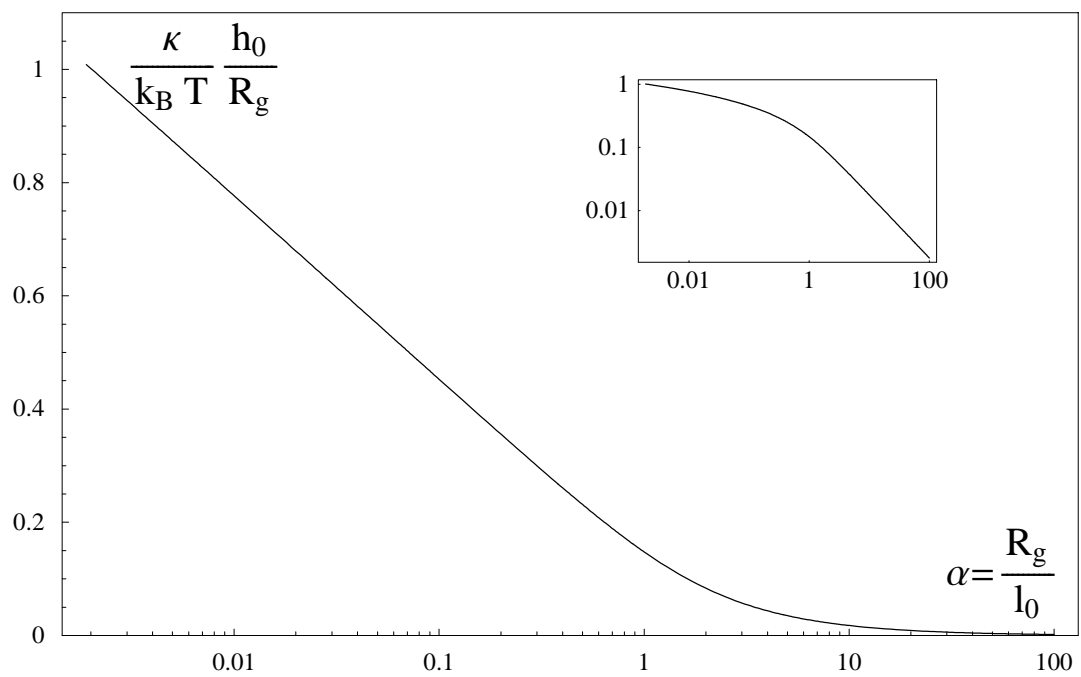
\end{document}